# In-air ion beam analysis with high spatial resolution proton microbeam

M. Jakšić, D. Chokheli, S. Fazinić, V. Grilj, N. Skukan, I. Sudić, T. Tadić and T. Antičić

Ruđer Bošković Institute, Bijenička cesta 54, 10000 Zagreb, Croatia

One of the possible ways to maintain the micrometer spatial resolution while performing ion beam analysis in the air is to increase the energy of ions. In order to explore capabilities and limitations of this approach, we have tested a range of proton beam energies (2 - 6 MeV) using in-air STIM (Scanning Ion Transmission Microscopy) setup. Measurements of the spatial resolution dependence on proton energy have been compared with SRIM simulation and modelling of proton multiple scattering by different approaches. Results were used to select experimental conditions in which 1 micrometer spatial resolution could be obtained. High resolution in-air microbeam could be applied for IBIC (Ion Beam Induced Charge) tests of large detectors used in nuclear and high energy physics that otherwise can not be tested in relatively small microbeam vacuum chambers.

1. Introduction

"In-air microbeam" or "external microbeam" refers to the ion microbeam of the MeV energy range extracted into the atmosphere. This can be performed either through a thin exit foil or a collimator. In order to reduce influence of the ion beam scattering and x-ray absorption, helium gas is often introduced into the volume between the exit foil and the sample. Collimating the ion beam to micrometer size by means of differentially pumped capillary or aperture provides another and cheap alternative to ion beam focusing using quadrupole multiplets [1-4]. Some authors used also thin exit foil for extraction of collimated ion beam into the atmosphere [5-6]. Disadvantage of collimated beam setup is necessity to scan the sample instead of a more versatile solution using the beam scanning. Another disadvantage could be also the small angle scattering of ions within the collimator [7] resulting in certain fraction of ions being non-monoenergetic [3].

The most of in-air microbeam configurations are therefore based on direct extraction of previously focused ion microbeam into the atmosphere through a thin exit foil. Protons in the energy range between 2.4 to 5 MeV are mainly used [8-22]. Since the influence of ion microbeam lateral spread on in-air microbeam spot size decreases with the decrease of working distance (distance between the exit foil and sample), samples are often placed on the exit foil itself [8-11].

Many authors reported experimental measurements and/or modelling of the beam spot size degradation as a function of working distance for different exit foil materials and its thicknesses. Reported working distances were from 100 μm to several mm [12-22]. T. Tadić et al. [12] estimated 96 μm in-air microbeam resolution for focused 4 MeV proton beam and a 50 μm thick polyimide foil, while the sample was placed in air, less than 100 μm from the exit foil. G.W. Grime et al. [14] used 3 MeV protons and beam path of 4 mm, with minimum spot size (FWHM) of 60 μm with Kapton exit foil and 15 μm using silicon nitride ($Si_3N_4$) film. M.Á. Ontalba Salamanca et al. [15] used 2.4 MeV proton microbeam, 8 μm Al exit foil and sample placed in air 3.5 mm from the foil. The incident beam energy at the sample surface was over 2.22 MeV and the beam spot size was about 120 μm. O. Enguita et al. [17] used 3 MeV protons, 8 μm Kapton exit foil to obtain beam spot sizes above 53 μm at working distances above 2 mm. K. Yasuda et al. [18] used 100 nm thick $Si_3N_4$ film as an exit foil for extraction of proton microbeam into the air or in mixture of helium gas and air. When the distance from exit foil was 0.2 mm, the beam size was reported to be only 4 μm in case of the proton energy of 2.5 MeV (air and helium gas) and 5 μm in case of the proton energy of 1.7 MeV (air). E. Colombo et al. [19] used 100 nm thick $Si_3N_4$ film and 3 MeV proton microbeam, achieving 10 μm beam spot on sample placed in the air 2.5 mm from the foil. T. Konishi et al. [20] used 1 μm thick $Si_3N_4$ film as an exit foil for extraction of 3.4 MeV proton microbeam in the air achieving 5 μm beam spot with 100 μm working distance in the air. One of very few reported in-air heavy ion microbeams was published by L. Sheng et. al [23], who used 200 nm $Si_3N_4$ film as an exit foil for extraction of $^{40}Ar^{15+}$ ions with an energy of 25 MeV/u, achieving 2 μm beam spot on 100 μm distance from the foil.

From the above discussion it is evident that the best in-air microbeam spatial resolution could be obtained with $Si_3N_4$ exit foil. It can be also concluded that only few facilities are working with beam spot sizes below 10 μm. As there are many application possibilities that require much better spatial resolution (e.g. ~ 1 μm), the main objective of this work is to explore possibilities to obtain micrometer spatial resolution. Quite obvious way to obtain better spatial resolution is the use of higher proton beam energy, but this would not always be the best choice, having in mind peculiarities of various ion beam analysis techniques and the vide range of possible applications.

## 2. Experimental

In order to introduce in-air microprobe possibility at the Ruđer Bošković Institute (RBI) microprobe facility, a nozzle with a microbeam exit foil has been added at the beam out flange of the spherical vacuum chamber [24]. As it is shown on Figure 1, sample for STIM or IBIC measurements is mounted on the miniature xyz goniometer stage. Owing to the fact that working distance of the RBI heavy ion microprobe is only 110 mm, while the distance from the center to the face of the exit flange is 160 mm, it has been possible to construct a beam exit nozzle with an exit foil placed at the 290 mm distance from the last quadrupole. Unfortunately such a large distance cannot be used to obtain high spatial resolution for conventional ion beam analysis techniques (e.g. PIXE) that require relatively high microbeam currents (pA to nA range). Still, by the significant reduction of object and collimator apertures, and low current mode (~fA range) operation suitable for techniques such as STIM and IBIC, spatial resolution down to 1 μm could be obtained in spite of the relatively long focus. Therefore, in order to study behaviour of the beam size degradation as a function of proton beam energy and distance from the exit foil, STIM technique has been employed.

Two types of exit foils have been used at the new in-air microbeam setup at RBI. The first one is $Si_3N_4$ foil today used almost as a standard and whose behaviour could serve as a reference. The second one is recently developed diamond membrane window. Its capabilities opened possibilities for various applications, in particular for irradiation using single ions [25].

Degradation of the beam spot was monitored for three proton energies (2, 4 and 6 MeV) and the 100 nm thick $Si_3N_4$ exit foil. After focussing the beam at the quartz disc, the current was reduced by closing the object and collimator slits. In this way low microbeam current (~ 1000 p/s) and low beam divergence (x mrad) has been obtained. Hamamatsu PIN diode S1223-01 with 500 mesh Ni grid placed on top of it was mounted on the precision Z stage that enabled fine adjustment of the exit foil to diode distance. By scanning the beam over the grid edge, a 2D STIM intensity distribution map was obtained. For each beam energy and working distance, several STIM profile measurements were fitted to the error function.

In order to be able to predict the behaviour of in-air microbeam as a function of proton energy, modelling of the process has been performed as well. The microbeam spatial resolution degradation was simulated as a function of exit foil material and beam path length in air by two approaches. The first conventional approach is using SRIM [26] simulation that is known to underestimate degradation of the microbeam spot size [27]. Therefore we have also used a model of small angle multiple scattering being developed by T. Tadić [12,13], on the basis of the most successful theory of small-angle multiple scattering of ions in the screened Coulomb potential, earlier proposed by P. Sigmund and K. B. Winterborn [28], as well as by A. D. Marwick and P. Sigmund [29].

3. **Results and discussion**

Experimental results are presented at Figure 2 for three proton energies, 100 nm $Si_3N_4$ exit foil and distances up to 10 mm from the exit foil. Due to the relatively long working distance of the in-air test setup (30 cm), initial microbeam size was 5 μm. As expected [27], experimental results of the microbeam resolution degradation were higher than the SRIM prediction, but contrary to SRIM our calculation using the modelling of small angle multiple scattering showed results that are much closer to the experimental ones. A possible explanation for this is difference between the screening lengths $a_{TF}$ for Thomas-Fermi screened potential and $a_{ZBL}$ Ziegler-Biersack-Littmark screening potential used in SRIM. Namely, although these potentials have nearly the same shape for radiuses smaller than five screening lengths [30], the Thomas-Fermi screening lengths $a_{TF}$ for protons are about two times larger than $a_{ZBL}$, i.e. $a_{TF} = 2.08\ a_{ZBL}$ for protons in diamond and $a_{TF} = 2.38\ a_{ZBL}$ for protons in $Si_3N_4$.

Simulations did not take into account possible microbeam divergence that could also influence degradation of the ion beam focus. In fact, the STIM setup being used in this work had very low beam divergence and therefore did not influence the results presented. In a high current mode, that has generally much higher beam divergence, this has to be taken into account as well. Contribution of beam divergence is generally independent on ion energy and therefore for higher proton energies it may have a larger effect on in-air microbeam resolution than the ion beam lateral spread. Therefore, for the high divergence microbeams, sample has to be positioned in the microbeam focal plane as close as possible. For example at the RBI microprobe setup operating in a high current mode, the ion beam waist within the last of the focusing quadrupoles is ~1 mm. For the microbeam working distance of 300 mm, the ion beam convergence angle at focal plane equals to 3.3 mrad, which results in the microbeam spot spread of ~3 μm at distance of 1 mm after the focus position.

Considering that multiple scattering modelling and to a certain extent SRIM simulation could give reliable prediction of the beam broadening, we have calculated influence of the exit foil and of the air path for several typical experimental conditions. Although it is shown before that SRIM underestimates beam broadening, data shown in Table 1 suggest that submicrometer spatial resolution in the air could be obtained for proton energies above 6 MeV, for $Si_3N_4$ exit foil and less than 0.5 mm working distance. The simulation was also performed for the diamond membrane exit foil, due to its possible application as a trigger for single ion irradiation [25]. In this case, proton beam broadening for energies above 6 MeV has been estimated to below 5 μm which may be useful for certain applications.

In order to demonstrate applicability of higher energy (>6 MeV) proton microbeams, we present here IBIC measurements performed on silicon power diode with a complex guard ring structure at its edges. In vacuum and in air measurements were performed and results are shown in Figure 3. This example shows capability to image complex detector structures with different responses coming from different regions of the exposed sample. Such conditions are typical for semiconductor detectors used in high-energy physics which are generally embedded into the complex structure that in addition to the detector diode itself, consists of several other layers placed on top of each other. These include readout electronics above the detector, network of connecting lines and different layers of insulators. Only higher energy protons (6 MeV has penetration depth in silicon of about 500 um) can penetrate subsequent layers above the detector in order to obtain the measurable IBIC signal. In addition, such detectors could be tested only outside the scattering chamber due to their large size.

**Conclusions**

By performing measurements that are confirmed by simulations, we have shown that an in-air Ion Beam Analysis setup with capability to reach 1 μm spot size could be designed if the higher energy protons (more than 6 MeV) are used along with a small (less than 0.5 mm) distance between the sample and the microprobe exit foil. This is in particular useful for the low current techniques such as STIM and IBIC. Modelling of small angle multiple scattering performed in this work showed acceptable agreement with an experiment, which allows us to assume that 1 μm spatial resolution could be obtained in certain cases using proton beam energy above 6 MeV. Although not discussed in this work, the use of helium atmosphere could improve the microbeam spatial resolution even more and allow use of even longer distances between the exit foil and the sample.

**Acknowledgement**

This work has been supported in part by the Croatian Science Foundation under the project 8127 (MIOBICC), the European Community under the FP7 project Particle detectors – EC contract no. 256783 and the IAEA Research Contract CRO-17051.

**References**

[1] K. Ishii, N. Fujita, H. Ogawa, Nucl. Instr. and Meth. B 269 (2011) 1026–1028

[2] Z-W. Hu, L-Y. Chen, J. Li, B. Chen, M-L. Xu, L. Qin, L-J. Wu, F-R Zhan, Z.-L. Yu, Nucl. Instr. and Meth. B 244 (2006) 462–466

[3] K. Yasuda, M. Hatashita, S. Hatori, T. Inomata, R. Ishigami, Y. Ito, T. Kurita, M. Sasase, K. Takagi, Nucl. Instr. and Meth. B 210 (2003) 27–32

[4] M.J. Simon, M. Döbeli, A.M. Müller, H.-A. Synal, Nucl. Instr. and Meth. B 273 (2012) 237–240

[5] H. Khodja, M. Hanot, M. Carriere, J. Hoarau, J.F. Angulo, Nucl. Instr. and Meth. B 267 (2009) 1999–2002

[6] N. Grassi, Nucl. Instr. and Meth. B 267 (2009) 825–831

[7] S. Incerti, Ph. Barberet, B. Courtois, C. Michelet-Habchi, Ph. Moretto, Nucl. Instr. and Meth. B 210 (2003) 92–97

[8] J. A. Cookson and F. D. Pilling, Phys. Med. Biol. 6 (1976) 965-969

[9] T. Sakai, Y. Naitoh, T. Kamiya, Y. Kobayashi, Nucl. Instr. and Meth. B 158 (1999) 250-254

[10] M. Oikawa, T. Kamiya, M. Fukuda, S. Okumura, H. Inoue, S. Masuno, S. Umemiya, Y. Oshiyama, Y. Taira, Nucl. Instr. and Meth. B 210 (2003) 54–58

[11] K-D. Greif, H. J. Brede, D. Frankenberg, U. Giesen, Nucl. Instr. and Meth. B 217 (2004) 505–512

[12] T. Tadić, M. Jakšić, C. Capiglia, Y. Saito, P. Mustarelli, Nucl. Instr. and Meth. B 161-163 (2000) 614-618

[13] T. Tadić, M. Jakšić, Z. Medunić, E. Quartarone, P. Mustarelli, Nucl. Instr. and Meth. B 181 (2001) 404-407

[14] G.W. Grime, M.H. Abraham, M.A. Marsh, Nucl. Instr. and Meth. B 181 (2001) 66-70

[15] M.Á. Ontalba Salamanca, F.J. Ager, M.D. Ynsa, B.M. Gómez Tubío, M.Á. Respaldiza, J. García López, F. Fernández-Gómez, M.L. de la Bandera, G.W. Grime, Nucl. Instr. and Meth. B 181 (2001) 664–669

[16] C. Remazeilles, V. Quilleta, Th. Calligaro, J. C. Dran, L. Pichon, J. Salomon, Nucl. Instr. and Meth. B 181 (2001) 681–687

[17] O. Enguita, M.T. Fernandez-Jimenez, G. Garcia, A. Climent-Font, T. Calderon, G.W. Grime, Nucl. Instr. and Meth. B 219–220 (2004) 384–388

[18] K. Yasuda, V.H. Hai, M. Nomachi, Y. Sugaya, H. Yamamoto, Nucl. Instr. and Meth. B 260 (2007) 207–212

[19] E. Colombo, S. Calusi, R. Cossio, L. Giuntini, A. Lo Giudice, P.A. Mando, C. Manfredotti, M. Massi, F.A. Mirto, E. Vittone, Nucl. Instr. and Meth. B 266 (2008) 1527–1532

[20] T. Konishi, T. Ishikawa, H. Iso, N. Yasuda, M. Oikawa, Y. Higuchi, T. Kato, K. Hafer, K. Kodama, T. Hamano, N. Suya, H. Imaseki, Nucl. Instr. and Meth. B 267 (2009) 2171–2175

[21] T. Calligaro, Y. Coquinot, L. Pichon, B. Moignard, Nucl. Instr. and Meth. B 269 (2011) 2364–2372

[22] K. Yasuda, M. Nomachi, Y. Sugaya, H. Yamamoto, H. Komatsu, Nucl. Instr. and Meth. B 269 (2011) 2180–2183

[23] L. Sheng, M. Song, X. Zhang, X. Yang, D. Gao, Y. He, B. Zhang, J. Liu, Y. Sun, B. Dang, W. Li, H. Su, K. Man, Y. Guo, Z. Wang, G. Xiao, Nucl. Instr. and Meth. B 269 (2011) 2189–2192

[24] M. Jakšić, I. Bogdanović Radović, M Bogovac, V. Desnica, S. Fazinić, M. Karlušić, Z. Medunić, H. Muto, Ž. Pastuović, Z. Siketić, N. Skukan, T. Tadić, Nucl. Instr. and Meth. B260 (2007) , 114-118

[25] V Grilj, N Skukan, M Pomorski, W Kada, N Iwamoto, T. Kamiya, T. Ohshima, M. Jakšić, Appl. Phys. Lett. 103 (2013) 243106

[26] J.F. Ziegler, J.P. Biersack (Eds.), SRIM, the Stopping and Range of Ions in Matter, © IBM Co 1998. The SRIM home page: http://www.srim.org/.

[27] C. Michelet, Ph. Moretto, G. Laurent, W.J. Przybylowicz, V.M. Prozesky, C.A. Pineda, Ph. Barberet, F. Lhoste, J. Kennedy, Nucl. Instr. and Meth. B 181 ( 2001) 157–163

[28] P. Sigmund and K. B. Winterborn, Nucl. Instr. and Meth. 119 (1974) 541–557

[29] A. D. Marwick and P. Sigmund, Nucl. Instr. and Meth. 126 (1975) 317–323

[30] W. Eckstein, "Computer Simulation of Ion-Solid Interactions", Springer Series in Materials Science, Vol. 10 (1991) 40-62

| Energy / air path | 100 nm Si$_3$N$_4$ | 6 μm diamond |
|---|---|---|
| 3 MeV / 0.5 mm | 1.02 | 9.0 |
| 3 MeV / 2.0 mm | 4.39 | 30.6 |
| 6 MeV / 0.5 mm | **0.50** | 4.3 |
| 6 MeV / 2.0 mm | 2.06 | 14.8 |
| 9 MeV / 0.5mm | **0.34** | 2.9 |
| 9 MeV / 2.0 mm | 1.40 | 9.9 |

Table 1. SRIM simulation of the beam focus broadening (FWHM in μm) after passage of proton microbeam through Si$_3$N$_4$ and diamond exit foils and 0.5mm and 2.0 mm air path.

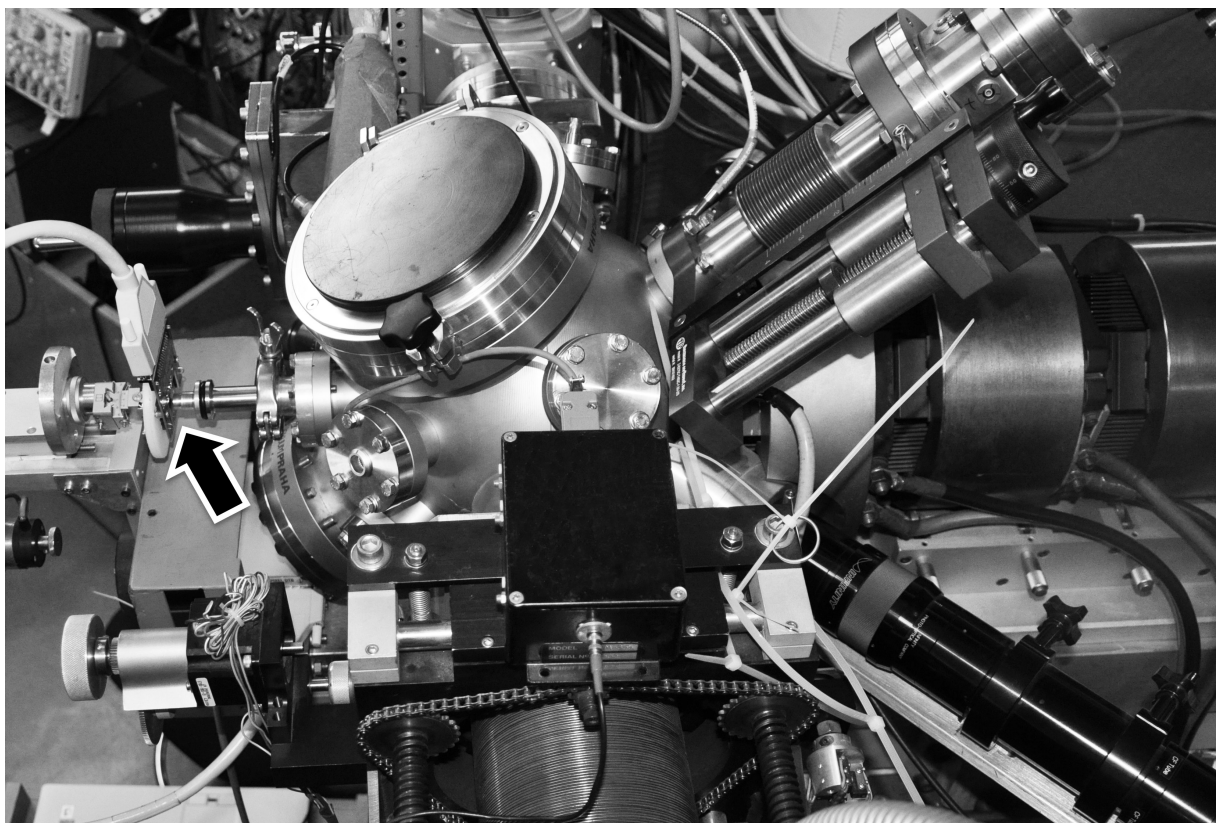

Figure 1. A side view of the RBI microprobe experimental setup. A quadrupole triplet lens is positioned on the right side of the spherical scattering chamber. Arrow (on left) indicates the position of the in-air microbeam focus with a 64 x 64 silicon pixel detector mounted for IBIC tests.

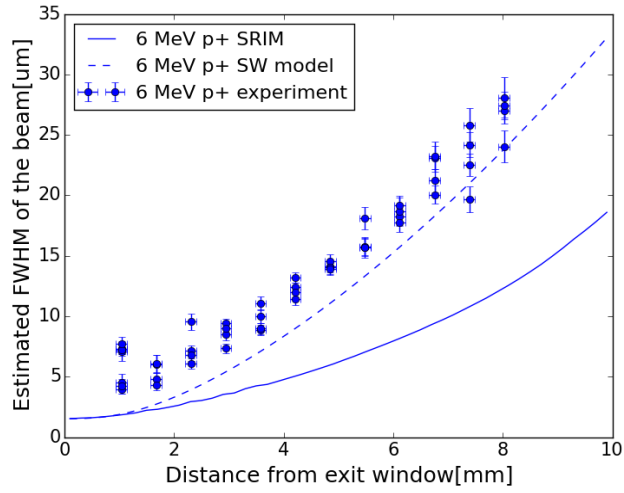

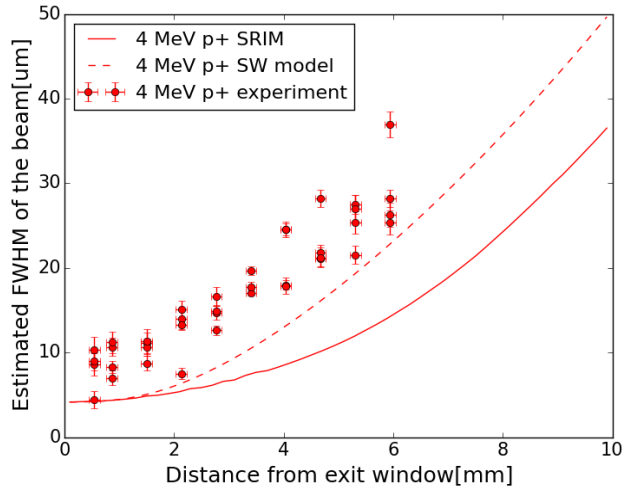

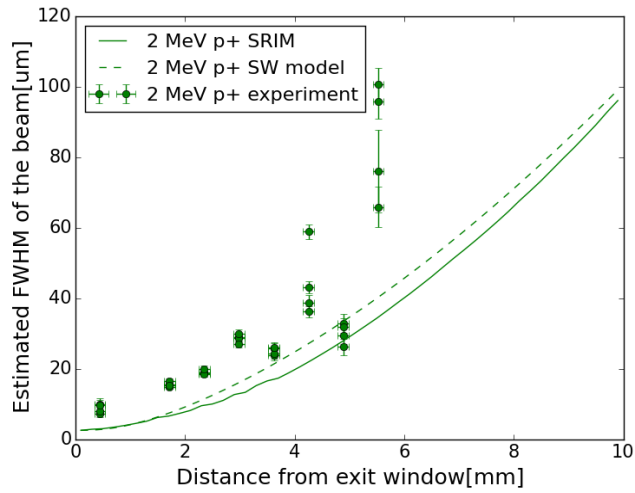

Figure 2. Experimental results, SRIM simulation (full line) and Sigmund and Winterborn model (dotted line) for microbeam broadening in air as a function of distance for the $Si_3N_4$ exit foil and three different proton energies ( 2, 4 and 6 MeV ).

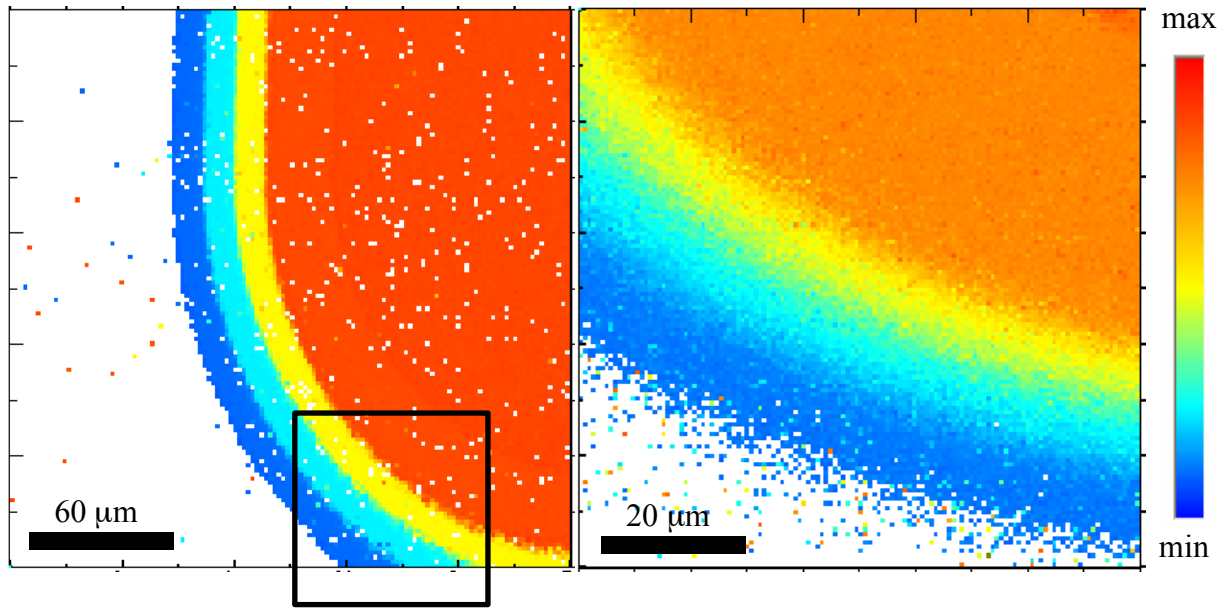

Figure 3. IBIC image of the edge of the Si power diode, left done in vacuum by 2 MeV protons, right done in air by 4 MeV protons after $Si_3N_4$ exit foil and 2 mm of air.